\documentclass[sigconf]{acmart}
\usepackage{xspace}
\usepackage[inline]{enumitem}
\usepackage{float}
\usepackage{tikz}
\usepackage{algpseudocode}
\usepackage{amsmath}
\usepackage[ruled,vlined,noend]{algorithm2e}
\usepackage{balance}
\usepackage{mathtools}
\usepackage{listings}

\SetCommentSty{mycommfont}

\SetKw{Break}{break}
\SetKwInOut{Input}{Input}
\SetKwInOut{Output}{Output}
\SetKwRepeat{Do}{do}{while}
\SetKw{Break}{break}

\newtheorem{example}{Example}

\AtBeginDocument{%
  }

\setcopyright{acmlicensed}
\copyrightyear{2026}
\acmYear{2026}
\acmDOI{XXXXXXX.XXXXXXX}
\acmConference[VLDB '26 Demos]{}{Boston, USA}

\newcommand\tool{\textsc{WN--Wrangle}\xspace}
\newcommand\powder{\textsc{POWDER}\xspace}
\newcommand{\stepCounter}[1]{{\Large\textcircled{{\small#1}}}}

\DeclareRobustCommand{\annotation}[1]{%
  \tikz[baseline=(char.base)]{
    \node[rectangle, fill=white, inner sep=1.2pt] (char)
    {\textcolor{black}{\textbf{\small #1}}};
  }%
}

\DeclareRobustCommand{\annotationfig}[1]{%
  \tikz[baseline=(char.base)]{
    \node[shape=circle, fill=lightgray, inner sep=1.2pt] (char)
    {\textcolor{black}{\textbf{\small #1}}};
  }%
}

\begin{document}

\title{\tool: Wireless Network Data Wrangling Assistant}

\author{Anirudh Kamath}
\affiliation{
  \institution{University of Utah}
  % \city{Salt Lake City}
  % \state{UT}
  \country{USA}}
\email{anirudh.kamath@utah.edu}

\author{Dustin Maas}
\affiliation{
  \institution{University of Utah}
  % \city{Salt Lake City}
  % \state{UT}
  \country{USA}}
\email{dmaas@cs.utah.edu}

\author{Jacobus Van der Merwe}
\affiliation{
  \institution{University of Utah}
  % \city{Salt Lake City}
  % \state{UT}
  \country{USA}}
\email{kobus@cs.utah.edu}

\author{Anna Fariha}
\affiliation{
  \institution{University of Utah}
  % \city{Salt Lake City}
  % \state{UT}
  \country{USA}}
\email{afariha@cs.utah.edu}

\settopmatter{printacmref=false} % Remove acm reference for submission
\settopmatter{printfolios=true} % add page numbers
\renewcommand\footnotetextcopyrightpermission[1]{} % Remove copyright block

\begin{abstract}

\looseness-1 Data wrangling continues to be the most time-consuming task in the
data science pipeline and wireless network data is no exception. Prior
approaches for automatic or assisted data-wrangling primarily target unordered,
single-table data. However, unlike traditional datasets where rows in a table
are unordered and assumed to be independent of each other, wireless network
datasets are often collected across multiple measurement devices, producing
\emph{multiple}, \emph{temporally ordered} tables that must be integrated for
obtaining the complete dataset. For instance, to create a dataset of the signal
quality of 5G cell towers within a geographic region, GPS data collected by
cellphones must be joined with radio-frequency measurements of the
corresponding cell towers. However, the join key \texttt{time\-stamp} typically
exhibits mismatched sampling periods, causing a \emph{misalignment}.
Data-wrangling techniques for generic time-series datasets also fail here,
since they lack knowledge of domain-specific data semantics---often defined by
network protocols and system configurations. To aid in wrangling wireless
network datasets, we demonstrate \tool, an \emph{interactive wrangling
assistant}---tailored to the wireless network domain---that \emph{suggests} the
top-k next-best wrangling operations, along with rich, domain-specific
\emph{explanations}. Under the hood, \tool enforces temporal-constraints- and
a wireless-network-semantics-aware mechanism to score and rank an extended set of
wrangling operators to improve the data quality. We demonstrate how \tool
identifies elusive data-quality issues specific to the wireless network domain
and suggests accurate wrangling steps over datasets obtained from the widely
used \powder city-scale wireless testbed.

\noindent {\small Link to demo video: \textcolor{blue}{\url{https://users.cs.utah.edu/~afariha/wnwrangle.mp4}}}
\vspace{-3mm}

% \af{I kind of finalized the abstract. Don't make direct
% edits, rather leave comments.}
% \af{I also simplified the title. It was too
% verbose. Stick to the main thing, which is DW assistance for WN data, right?
% You don't need to give everything in the title, rather something that makes it
% clear what this paper is all about.}

% \af{One thing missing in this abstract is How the tool works, what's the
% smarts or novel thing about the tool that is nontrivial to build as a simple
% engineer. Why is this tool not just a new feature of a software, rather a
% research prototype? What is your key idea that is new and not used before in
% other works?}

\end{abstract}

\maketitle

% \ifdefempty{\vldbavailabilityurl}{}{
% \vspace{.3cm}
% \begingroup\small\noindent\raggedright\textbf{PVLDB Artifact Availability:}\\
% The source code, data, and/or other artifacts have been made available at \url{\vldbavailabilityurl}.
% \endgroup

%!TEX root = main.tex

\section{Introduction}

The fifth generation (5G) of mobile networks is expected to serve nearly 3
billion users worldwide by 2026, with over 65\% of fixed wireless access
connections projected to be provided via 5G by
2025~\cite{ericsson_mobility_report25}. Supporting this scale demands
intelligent service monitoring, provisioning, and network planning, all of
which increasingly rely on data-driven AI/ML techniques. For instance, Vodafone
reported reducing data-ingestion latency from 36 hours to 25 minutes by
leveraging insights from 70 petabytes of user-collected
data~\cite{Peermamode2022VodafoneGoogleCloud}. However, a critical prerequisite
for learning from such data is making it \emph{analysis-ready}, which in
practice requires extensive and routine \emph{data wrangling} of large-scale
wireless network (WN) datasets. Data wrangling, in general, is often a tedious
and time-consuming process. Data scientists reportedly spend up to 60\% of
their time on tasks such as cleaning, imputing, transforming, and organizing
data~\cite{cowrangler}. This has motivated the development of
\emph{assistant tools} for data wrangling (DW), such as
CoWrangler~\cite{cowrangler} and Wrangler~\cite{wrangler}, which can suggest
relevant operations to expedite the DW process.

However, such general-purpose DW assistants~\cite{cowrangler, auto-suggest}
fall short
when applied to WN datasets for several reasons. \textbf{First}, they assume
that rows within a dataset are unordered and unrelated, often suggesting incorrect imputation or even dropping rows with missing values. For instance, CoWrangler's~\cite{cowrangler} imputation suggestions are limited to constant, mean, mode, and median. WN datasets,
however, are inherently temporal, with measurements recorded sequentially over time and often at fixed sampling rates. Na\"ively applying a mean imputation can distort
periodic signals; instead imputations must be done using appropriate forward or backward filling. \textbf{Second}, 
generic DW tools typically
operate on a single table, which fails when multiple tables must be joined via numerical join keys that are misaligned due to different periodicity or measurement units.
% ~\cite{deepsense}
For instance, Auto-Suggest~\cite{auto-suggest} fails to join over numerical keys with mismatched units (e.g., miles vs.\ kilometers).
% tables containing measurements at different periods need to be joined
% ---a common occurrence when preparing wireless datasets~\cite{deepsense}.
% on a
% single table at a time, which can fail catastrophically when multiple tables
% need to be joined, especially if they are misaligned~\cite{deepsense} due to
% different measurement periods. 
\textbf{Third}, existing DW assistants are
domain-agnostic and ignore crucial semantics of WN data. For instance,
logarithmic units such as decibel-milliwatts must be converted to linear units before
aggregation. We proceed to highlight these issues in Example~\ref{ex:one}.

\begin{figure*}[t]

\resizebox{0.86\textwidth}{!}{\includegraphics[]{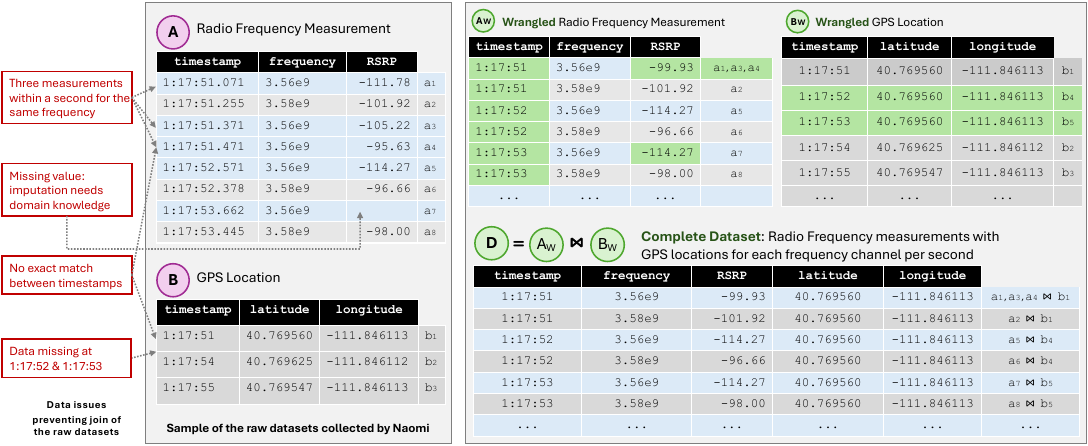}}

\vspace{-4mm} \caption{\small \annotationfig{A}: RF measurement data sample,
\annotationfig{B}: GPS data sample, \annotationfig{$\text{A}_\text{w}$} \&
\annotationfig{$\text{B}_\text{w}$}: desired wrangled datasets,
\annotationfig{D}: desired complete dataset.} \vspace{-3mm}

\label{fig:data_snippet}
\end{figure*}

\begin{example} 
\label{ex:one}
Naomi is collecting data to train an intelligent network configuration
management system (Figure~\ref{fig:data_snippet}). This system requires
per-second, complete radio frequency (RF) measurements across all frequency
channels, with corresponding GPS coordinates. She uses two devices for data
collection: an RF measurement device that records data in the format
\verb|<timestamp, frequency, RSRP>|\footnote{RSRP (Reference Signal Received
Power) measures the power of the reference signal received by a device from a
cell tower, typically expressed in dBm, a logarithmic unit.} as shown in
\annotation{A},
and a smartphone that records data in the format
\verb|<timestamp, latitude,| \verb|longitude>| 
as shown in \annotation{B}. Her goal is to generate a dataset in the schema
\verb|<timestamp, frequency, RSRP,| \verb|latitude, longitude>|
(shown in \annotation{D}) by joining~\annotation{A} and~\annotation{B} over the
join key \verb|timestamp|.

However, while trying to join \annotation{A} and~\annotation{B}, Naomi observes
that the measurements are logged at different temporal granularities:
\annotation{A} contains multiple entries within a second while \annotation{B}
has missing entries for some seconds. Furthermore, there is no exact match
between the \verb|timestamp| attributes of the two tables and Naomi must temporally align them. Existing tools like CoWrangler~\cite{cowrangler} overlooks Naomi's goal of joining \annotation{A} and \annotation{B}, and exacerbates the situation by making incorrect suggestions:
(1)~Dropping the
row $a_7$ from \annotation{A} due to the missing \texttt{\small RSRP}, which
results in losing the only entry for the corresponding timestamp and frequency.
(2)~Imputing $\mathtt{a_7[RSRP]}$ with the arithmetic mean of \texttt{\small RSRP}, despite the unit for \texttt{\small RSRP} being logarithmic
(decibel-milliwatts).

To join \annotation{A} and \annotation{B}, Naomi must ensure that (1)~each
\texttt{\small frequency} channel has exactly one \texttt{\small RSRP} reading
per second, as in \annotation{$\text{A}_\text{w}$}; (2)~exactly one
geo-location entry exists per second in the GPS data, as in
\annotation{$\text{B}_\text{w}$}; and (3)~the values in \texttt{\small
timestamp} in \annotation{A} \& \annotation{B} match exactly. To achieve this,
the following wrangling steps are essential:

\noindent \textbf{Step 1}: Forward-fill the missing \texttt{\small RSRP} in
$\mathtt{a_7}$ using the value from $\mathtt{a_5}$, but not $\mathtt{a_6}$
since the frequency channels are different.

\noindent \textbf{Step 2:} (a)~Merge rows $\mathtt{a_1}$, $\mathtt{a_3}$, and
$\mathtt{a_4}$ using techniques appropriate for logarithmic units---which
ensures a per-second periodicity in \annotation{$\text{A}_\text{w}$}---and
(b)~round down timestamps to the nearest second.

\noindent \textbf{Step 3:} (a)~Insert two empty rows between $\mathtt{b_1}$ and
$\mathtt{b_2}$ to fill the missing seconds and achieve a per-second periodicity
in \annotation{$\text{B}_\text{w}$}, followed by (b)~forward-filling location
values in the newly inserted rows, because smartphones suspend GPS logging
while stationary to conserve power.

Naomi had to spend 20 minutes writing and validating the mundane code snippet below---a disproportionate time cost despite her domain expertise---to obtain the desired dataset \annotation{D} = \annotation{$\text{A}\text{w}$} $\bowtie$ \annotation{$\text{B}\text{w}$}.
\vspace{1mm}
% \vspace{-4mm}
% \subsubsection*{\underline{The issue of re-alignment for joins is ubiquitous}} \looseness-1
% Timestamped datasets often require re-alignment for joins, as observed in domains such as climate science~\cite{morrison_wind_profile_dataset} and satellite networks~\cite{lottermoser2026measuringweathereffectslink}. 
% This need extends beyond temporal alignment in WN data, where multiple abstraction layers define column semantics. 
% For example, frequency channels (MHz) are also represented as (E)ARFCN,\footnote{(E)ARFCN stands for (E-UTRA) Absolute Radio Frequency Channel Number} a 3GPP-standardized identifier for 4G/5G channels~\cite{3gpp.36.101}. 
% Thus, integrating measurement and protocol-level data requires transformations (e.g., MHz to (E)ARFCN)~\cite{wnwrangle_tr}.

\definecolor{codebg}{gray}{0.94}

\lstset{
  language=Python,
  basicstyle=\scriptsize\ttfamily,
  breaklines=true,
  backgroundcolor=\color{codebg},
  frame=single,
  rulecolor=\color{gray!50},
  xleftmargin=2pt,
  xrightmargin=2pt,
  aboveskip=0pt,
  belowskip=0pt,
  showstringspaces=false,
  tabsize=2,
  keywordstyle=\color{blue},
  commentstyle=\color{gray!70},
  stringstyle=\color{green!50!black}
}

\begin{lstlisting}
# (Step 1) Forward-fill RSRP per frequency
A["RSRP"] = A.groupby("frequency")["RSRP"].ffill()

# (Step 2) Per-second log-aware RF aggregation
A["sec"] = A["timestamp"].dt.floor("S")
A["RSRP_lin"] = 10**(A["RSRP"]/10)
A = A.groupby(["sec", "frequency"]).agg({"RSRP_lin" : "mean"})
A["RSRP"] = 10*A["RSRP_lin"].apply(log10)
A.drop(columns = "RSRP_lin", inplace=True)

# (Step 3) Per-second GPS forward-fill
B = B.set_index("timestamp").asfreq("1S").ffill().reset_index()
\end{lstlisting}
\end{example}
\vspace{-4mm}
% \subsubsection*{\underline{Specialized wrangling is required for time-series data}}
% \looseness-1 While Example~\ref{ex:one} illustrates 

\subsubsection*{\underline{A wrangling assistant for wireless network data}}
\looseness-1 Example~\ref{ex:one} demonstrates several wrangling operations
that are specific to WN domain and are typically overlooked by generic
wrangling assistants. This observation motivates the need for a specialized WN
wrangling assistant that can proactively \emph{suggest} domain-relevant
wrangling operations. Such an assistant must satisfy the following
requirements: (i)~respect temporal constraints, such as ensuring periodicity
and completeness (Step~1, Step~2~(a), \&~Step 3); (ii)~support automatic
alignment across tables through downsampling (Step~2~(a)), upsampling
(Step~3~(a)), and homogenization (casting timestamp to the nearest second in
Step~2~(b)); (iii)~exploit inter-row relationships, including appropriate
imputation strategies (forward-filling in Step~1 and~Step~3~(b)); and
(iv)~preserve domain-specific semantic correctness, such as using imputation
methods suitable for logarithmic units (Step~2~(a)). Finally, to encourage
broad adoption, a WN wrangling assistant should additionally provide (v)~rich
\emph{explanations} for its suggested operations; and (vi)~\emph{interactive}
controls that allow users to inspect, customize, and guide the wrangling
process.

\subsubsection*{\underline{Temporal re-alignment is common across domains.}} Datasets containing temporal columns (i.e., containing timestamped measurements) inadvertently require temporal re-alignment to make them joinable. This is explicitly mentioned in multiple papers describing datasets in the domains of climate science~\cite{morrison_wind_profile_dataset}, energy consumption~\cite{pullinger2021ideal}, satellite networks~\cite{lottermoser2026measuringweathereffectslink}, and more.

\subsubsection*{\underline{Wrangling WN data does not mean only temporal re-alignment}}

In\\ WN data, there are multiple levels of abstraction that envelope the semantic meaning of columns. For example, frequency channels (often in the unit Mega-Hertz (MHz)) are represented as \emph{channel numbers} (termed EARFCN/ARFCN~\footnote{(E)ARFCN stands for (E-UTRA) Absolute Radio Frequency Channel Number}), which are 3GPP standardized representations for 4G/5G channels~\cite{3gpp.36.101}. Due to commercial measuring equipment often measuring channels in MHz, combining measurement data with protocol level datasets requires transformation from MHz to EARFCN/ARFCN. 
% We provide an illustrative example of such in our technical report~\cite{wnwrangle_tr}. However, we focus this demonstration on the common case of temporal re-alignment for data joins.

\smallskip \noindent While Example~\ref{ex:one} demonstrates wrangling operations
that are centered around timestamp re-alignment of wireless measurements, other entities (e.g., wireless channel representations) also require unit and format re-alignment, which we highlight in the example below.

\begin{figure*}[t]

\resizebox{0.86\textwidth}{!}{\includegraphics[]{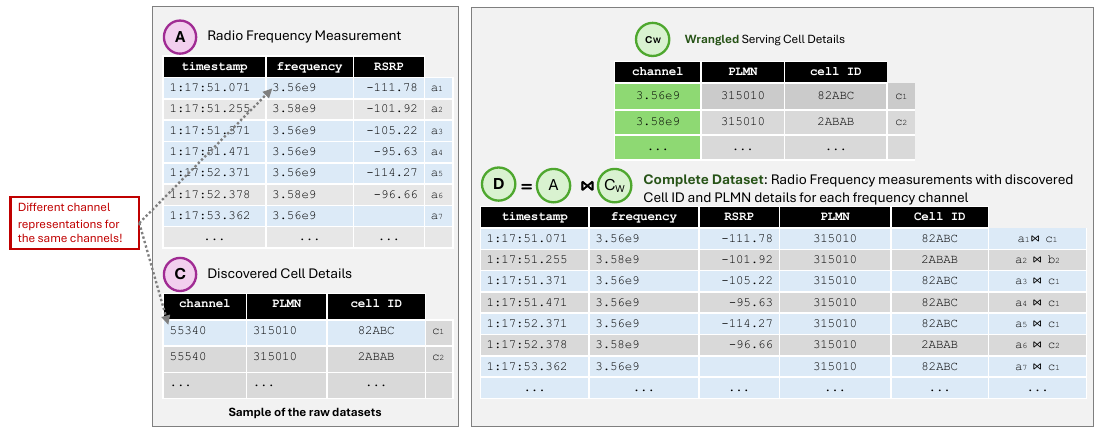}}

\vspace{-4mm} \caption{\small \annotationfig{A}: RF measurement data sample,
\annotationfig{C}: Discovered cells data sample, \annotationfig{$\text{c}_\text{w}$}: desired wrangled discovered cells dataset (with \texttt{\small channel} attribute converted to MHz),
\annotationfig{D}: desired complete dataset.} \vspace{-3mm}

\label{fig:ex2_data_snippet}
\end{figure*}

\begin{example} 
\label{ex:two}
Naomi now wishes to augment the RF measurements (\annotation{A} in Figure~\ref{fig:data_snippet}) with details about the 4G/5G cells in the area that were transmitting on these channels (Figure~\ref{fig:ex2_data_snippet}). For this, she procures an older dataset measured using a cellular modem device. This dataset contains the scanned cell towers in the form \verb|<timestamp,| \verb| channel_number,| \verb|cell_ID>|. In this scenario, Naomi wishes to join the two datasets by \verb|frequency| and \verb|channel_number| respectively, as they both contain the detected frequency that the cell tower transmitted on-- however, channel numbers are represented in EARFCN~\footnote{EARFCN stands for E-UTRA Absolute Radio Frequency Channel Number, an identifier for the center frequency of wireless transmissions by a radio device.}, whereas frequency values are represented in Megahertz (MHz). These are two units for the same entity, but she needs to map channel numbers to downlink frequency values in MHz using a translation table defined by a standard document (3GPP TS 36.101~\cite{3gpp.36.101}), where each channel number has different coefficients in the translation formula. For this, she must know each channel number that is represented in the RF measurements table, translate them separately, and then join the tables.
\end{example}

\smallskip \noindent This also extends to other WN entities that can be represented in different ways. For example global cell identifiers can be represented in hexadecimal or in decimal, requiring conversion from unit to another.

% \subsubsection*{\underline{Time-series wrangling across domains.}}
% \looseness-1 While Examples~\ref{ex:one} and~\ref{ex:two} demonstrate wrangling wireless datasets with a temporal component,
% such time-series data wrangling 

We demonstrate \tool, an interactive \textbf{W}ireless
\textbf{N}et\-work data-\textbf{Wrangl}ing assistant that suggests top-k
relevant wrangling operations in a multi-table setting, tailored for WN
datasets, satisfying the above six requirements. The key idea behind \tool is
\emph{temporal-constraint}-aware scoring of \emph{WN-specific} wrangling
operations, which builds on the notion of temporal functional
dependencies~\cite{temporal_rules_discovery_web_data_cleaning} to identify
periodicity violations in the data.

\subsubsection*{\underline{Related work}} \looseness-1 Prior works in the
databases community~\cite{cowrangler,wrangler} mainly target unordered
relational datasets and thus fail on WN datasets, which are temporal in nature.
Tools for cleaning temporal data using temporal integrity
constraints~\cite{temporal_rules_discovery_web_data_cleaning} assume that
domain-specific rules can be discovered from the data itself, which does not
hold for the WN domain, where rules are often dictated by network protocols or
system configurations. Recent works on time-series data
wrangling~\cite{autodwts} are domain-agnostic and fail to provide
alignment-related wrangling suggestions such as upsampling, downsampling, or
homogenizing (as shown in Example~\ref{ex:one}). While wrangling code
generation systems~\cite{wrangling_codegen_ase24} and foundation models have
achieved partial success due to their ability to understand data semantics,
they still struggle with complex tasks such as joining multiple incomplete
datasets with misaligned join keys. For instance, when asked to join the
datasets \annotation{A} and \annotation{B} while ensuring one record per
second, ChatGPT made several mistakes, including incorrect \texttt{\small RSRP}
imputation, incorrect frequency aggregation, and failing to
up/downsample~\cite{chatgpt_gps_wireless_data_join}. Commercial tools for
time-series data~\cite{datarobot} primarily target enterprise data sources and
require extensive manual configuration.

\smallskip\noindent\emph{\underline{Demonstration.}} \looseness-1 In our
demonstration, participants will observe how \tool identifies temporal
inconsistencies in two real-world datasets collected using the
POWDER~\cite{powder} city-scale wireless testbed, and suggests accurate
wrangling steps in an explainable and interactive manner. We will showcase how
a user can customize the suggestions in two WN data analysis scenarios---one to
improve the readability of radio-frequency readings on a map interface, and the
other to improve an ML model's performance on the collected data.

\smallskip

We provide an overview of \tool's inner working in Section 2, and a walkthrough
of the demonstration scenario based on Example~\ref{ex:one} in Section 3.

%!TEX root = main.tex

\section{System overview}
\label{sol_overview}

\tool targets multiple WN tables that share a context that allows their joint
use via joins 
% \ak{Specified that we support only joins}. 
While we focus on the temporal context in this
% demonstration, 
\tool also supports joining over frequency channels, and
spatial coordinates, and cell tower identities by leveraging the corresponding alignment semantics.

\smallskip\noindent \emph{\underline{Challenges.}} We identify the following
key challenges towards building \tool:
\textbf{(C1)}~How to \emph{model WN goal-oriented specific constraints} and \emph{detect
their violations} in the data?
\textbf{(C2)}~Which WN-specific wrangling operations should be considered as
\emph{candidates} for repairing the constraint violations?
\textbf{(C3)}~How to quantify the \emph{effectiveness} of candidate wrangling
operations---i.e., score and rank them---to enable accurate top-$k$ suggestions
automatically?
\textbf{(C4)}~How to generate domain-aware \emph{explanations} for the
suggestions, while allowing users to (optionally) guide the wrangling process
\emph{interactively}?

\begin{figure}[ht]
\resizebox{0.4\textwidth}{0.5\textwidth}{\includegraphics[]{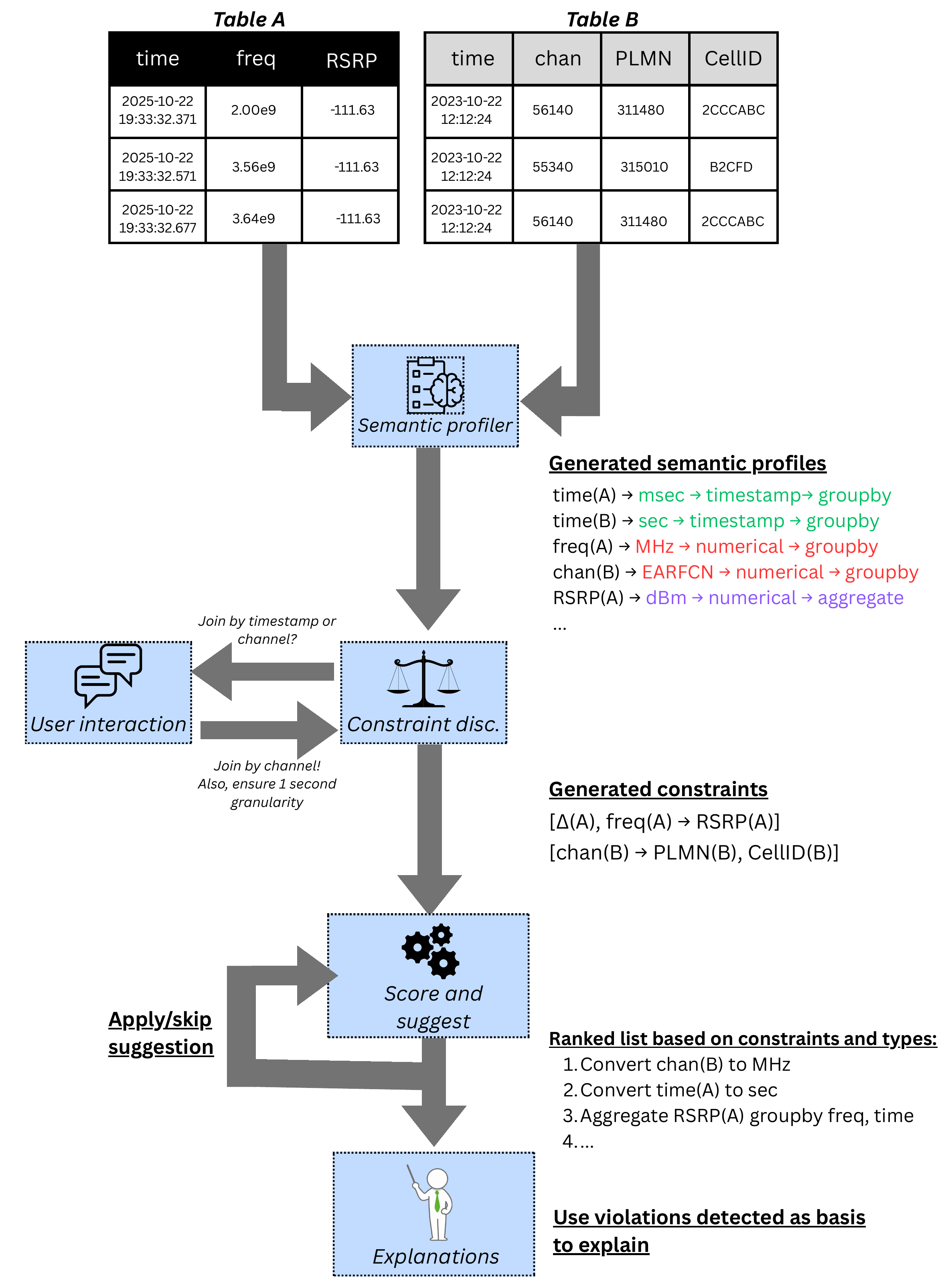}}

\vspace{-4mm} \caption{\small System architecture of \tool. After data upload, semantic profiling (Section~\ref{semantic_profiler}) generate feature profiles per column (units, type, aggregation/imputation role), the constraint discovery module (Section~\ref{rule_discovery}) uses these profiles to identify joinable columns and devises constraints to enforce to enhance the join, after which the scoring and suggestion of wrangling operations (Section~\ref{soloverview_ranking}) on the basis of constraints and feature profiles, and explanation synthesis (Section~\ref{explanation_module}) is done in a human-in-the-loop workflow, where the  user can apply suggestions or skip them over. 
% \ak{Allow users to define feature units, and edit aggregation and imputation strategies...}
} \vspace{-3mm}

\label{fig:sys_arch}
\end{figure}

\smallskip\noindent \emph{\underline{Overview.}} To address these challenges,
\tool comprises five components: (\S\ref{semantic_profiler})~a \emph{semantic
profiler} that profiles the data attributes; (\S\ref{rule_discovery})~a
\emph{constraint discovery module} that discovers WN-specific constraints;
(\S\ref{dsl})~a \emph{Domain Specific Language (DSL)} tailored towards WN data;
(\S\ref{soloverview_ranking})~a \emph{scoring method} that evaluates
effectiveness of candidate wrangling operations; and
(\S\ref{explanation_module})~an \emph{explanation module} that generates
explanations for the suggestions. The system architecture worklow is described in Figure~\ref{fig:sys_arch}.

\renewcommand{\thesubsubsection}{\thesection.\arabic{subsubsection}}

\subsubsection{\textbf{Semantic profiler}}
\label{semantic_profiler}
\tool must understand the data characteristics to model WN-specific constraints
(\textbf{C1}), support scoring of wrangling operations (\textbf{C3}), and
provide explanations (\textbf{C4}). To this end, \tool employs a \emph{semantic
profiler} that analyzes each data attribute over a small sample to infer the
following key aspects about them:

\begin{itemize}[leftmargin=*, topsep=0pt, partopsep=0pt, parsep=0pt, itemsep=0pt]

     \item its data type (e.g., numerical, categorical, ordinal);
	
     \item its semantic type, measurement unit, and scale (e.g., \texttt{\small
     RSRP} measures signal power in \texttt{\small dBm}, a logarithmic scale);
    
	 \item semantically correct aggregation strategies (e.g., using Frequency as a grouping column when aggregating RSRP);
\end{itemize}

\smallskip\noindent To obtain the above information, \tool queries a general-purpose LLM (GPT-5.2).
% ,eliminating the need for manual input from a domain expert. 
% While a
% general-purpose LLM suffices in most cases, a fine-tuned variant trained on
% WN-specific documents and manuals can further improve the semantic profiler.
These queries are issued carefully using crafted prompts to the LLM to (\textbf{Q1}) the column unit (e.g., dBm, or MHz), (\textbf{Q2}) gather the broad type (e.g., numerical or categorical) and (\textbf{Q3}) whether this column is ``grouping'' or ``aggregating'' (aggregating/imputation roles).
% Each unit supported by \tool (e.g., log(dBm)) is mapped to a pre-defined class (e.g., Signal Power).
% We synthesize such classes using the derived entity types from column descriptions generated with AutoDDG

\smallskip\noindent Below is the query template \tool crafts to assess the measurement units of values in the column (\textbf{Q1}). \tool's measurement units are pre-defined for wireless types. However, our demo also allows participants to configure their own units as keywords to ensure generalization when using datasets from other domains.

\begin{tikzpicture}
\node[
    draw,
    rounded corners,
    fill=pink!10,
    text width=8cm,
    align=left
] 
{\textbf{Question}:\\
You are an expert wireless data profiler.\\
I have a column in a pandas dataframe named chan. \\
It has a minimum value of 800.0, max value of 68661.0, and 18 unique values.\\
Here are some sample values: \\
\dots \\
Tell me the unit of this column's values- only choose from ['no unit', 'MHz', 'Hz', 'EARFCN', 'dBm', 'dB', 'hexadecimal', 'decimal', 'seconds', 'milliseconds'].\\ 
For example, a test score has no unit, so answer "no unit". \\ 
However, a temperature has a unit "celsius". \\
Answer only either ['no unit', 'MHz', 'Hz', 'EARFCN', 'dBm', 'dB', 'hexadecimal', 'decimal'], and nothing else. \\
\textbf{Answer}:\\
EARFCN
};
\end{tikzpicture}

After this, \tool assesses the broad data type (\textbf{Q2}) for each unit via the query template shown below. To do this, \tool uses the result of the previous query, and asks the LLM if data from a column labeled with a certain unit should be numerical, categorical, or ordinal. For example, the \texttt{\small RSRP} column is profiled with a unit of \texttt{\small dBm}, so it is ``numerical'' and should be treated as such. 

\usetikzlibrary{shapes.callouts}
\begin{tikzpicture}
\node[
    draw,
    rounded corners,
    fill=pink!10,
    text width=8cm,
    align=left
] 
{\textbf{Question}:\\
You are an expert wireless data profiler.\\
I have a data in a column named RSRP with the unit of dBm.\\
Here is a sample of that data: \\
\dots \\
Tell me if data with this unit's values are timestamp, categorical, numerical, or ordinal. \\
For example, temperature's unit is celsius, so it is a "number", even if some values are string values in it (to represent missing values).\\
For example, even though a bank account ID has no unit, it is "categorical" because the records can be grouped by bank account ID.\\
For example, even though a score ranging from 1 to 5 is a number, because it has no unit, it is "ordinal".\\
Answer only either "timestamp", "categorical", "numerical", or "ordinal", and nothing else.
    
\textbf{Answer}:\\
numerical
};

\end{tikzpicture}

\smallskip\noindent Finally, \tool gathers information about per-column aggregation and imputation roles (\textbf{Q3}). The LLM is asked if the context of the attribute's values warrants its usage as a ``grouping'' attribute (e.g., if it is categorical), or if it is an ``aggregating'' attribute (e.g., it is a numerical value changing over time). This is crucial, because even though some columns may be profiled as numerical, it may well be a column that must be used to group together rows in aggregations-- for example, the \texttt{\small freq} column in Table~\annotation{A} (Figure~\ref{fig:sys_arch}) is numerical, but it is definitely a grouping attribute when aggregating \texttt{\small RSRP} values to get a per-second granularity.

\begin{tikzpicture}
\node[
    draw,
    rounded corners,
    fill=pink!10,
    text width=8cm,
    align=left
] 
{\textbf{Question}:\\
You are an expert wireless data profiler. \\
I have a column in a pandas dataframe named chan, with other columns in the dataframe being ['cellid', 'PLMN'].\\
It has a unit of EARFCN.\\
Here are some sample values: \\
\dots \\
You should decide if this column goes in the `agg()` part ("aggregating"),\\
or if this column goes in the `groupby` part ("grouping")\\
of an aggregation query in pandas.\\
For example, in a dataframe with columns <date,product\_id,units\_sold>, date and product\_id are assigned "grouping", and units\_sold "aggregating".\\
Answer only either "aggregating" or "grouping" and nothing else.\\
\textbf{Answer}:\\
grouping
};
\end{tikzpicture}
% If the LLM profiles a column to be ``aggregating'', then an additional query is made to assess what function (from a pre-defined list) to use to aggregate values of this column.

% \begin{tikzpicture}
% \node[
%     draw,
%     rounded corners,
%     fill=pink!10,
%     text width=8cm,
%     align=left
% ] 
% {\textbf{Question}:\\
% You are an expert wireless data profiler. \\
% I have a column in a pandas dataframe named RSRP. \\
% It has a unit of dBm. \\
% Here are some sample values:\\
% \dots \\
% You should decide what aggregation method to use to combine multiple values into one data point.\\
% Your choices are: ['mean', 'sum', 'mode', 'median', 'log\_mean'].  \\
% Their descriptions are: ['arithmetic mean', 'sum', 'most frequent value', '50th percentile value', 'convert to linear domain, do arithmetic mean, convert back to log domain']. \\
% Answer only one of ['mean', 'sum', 'mode', 'median', 'log\_mean'] and nothing else.\\
% \textbf{Answer}:\\
% log\_mean
% };
% \end{tikzpicture}

\smallskip\noindent Using the information gathered from the LLM responses, \tool synthesizes the following based on the attribute's profile:
\begin{enumerate}
	 \item semantic-type- and domain-aware imputation strategies (e.g., forward-filling GPS locations, as columns with data type \texttt{\small decimal degrees} unit is imputed by forward filling); and

  \item semantic-type- and domain-aware aggregation strategies (e.g., converting to linear scale before averaging logarithmic values, as \texttt{\small dBm} unit values require this special technique for accurate aggregation); and
    
	 \item domain-aware type casting between attributes (e.g., \texttt{\small chan} attribute in Table \annotation{B} in Figure~\ref{fig:sys_arch} should be converted to MHz, the unit of \texttt{\small freq} channel).

\end{enumerate}

\smallskip \noindent \emph{Remark:} Our demo only supports type casting (item 3 above) between pre-configured wireless data types (e.g., EARFCN to MHz), as this requires explicit programming of a casting function in the DSL (Section~\ref{dsl}). Adding more type casting functions to the DSL is trivial per unit pair. 
% We can have users program their own type casting functions to add to the DSL if needed.

\subsubsection{\textbf{Constraint discovery module}} 
\label{rule_discovery}
To address \textbf{C1}, this module models WN-specific data constraints (e.g.,
temporal constraints requiring per-second measurements) and detects their
violations in the data. Since enforcing these constraints ensures temporal and
informational completeness, they provide guidance for suggesting wrangling
operations that reduce constraint violations. To model temporal constraints, we
use the established notion of temporal functional dependencies
(TFDs)~\cite{temporal_rules_discovery_web_data_cleaning}. For example, $TFD_1 =
[\Delta,$ \texttt{\small frequency}$] {\rightarrow}$ \texttt{\small RSRP}
denotes the temporal constraint of having ``exactly one \texttt{\small RSRP}
record per second per \texttt{\small frequency} channel'', where
$\Delta=\texttt{\small 1s}$ denotes a fixed periodicity in \texttt{\small
timestamp}.

\smallskip \noindent \emph{User interaction.} The constraint discovery module first confirms with the user which columns to use to join the tables. It does so by presenting potential options based on similar units which can be converted (e.g., milliseconds to seconds), or transformed (e.g., EARFCN to MHz). For instance, in Figure~\ref{fig:sys_arch}, the user does not wish to join over timestamps, and instead, the constraint discovery module recognizes (from semantic profiles) that EARFCN and MHz are units representing transferable units that can be transformed to each other.

\smallskip \noindent\emph{{Synthesizing constraints.}} Once the user confirms the joining columns, \tool synthesizes
TFDs and associated parameters by leveraging semantic data profiles---which
provides derived knowledge such as ``within a second granularity in the RF measurements table, it is required that each frequency (grouping categorical attribute) requires non-empty \texttt{\small RSRP}'' (to ensure no loss of temporal granularity).
% are typically recorded at a
% per-second granularity
% ---and by analyzing the temporal periodicities in the
% data tables under consideration. 
Specifically, \tool identifies a common
periodicity $\Delta$ that can be achieved across the data tables without
impacting too many tuples (minimizing side-effects) and confirms this with the user. While TFD
discovery~\cite{temporal_rules_discovery_web_data_cleaning} is fully automated,
\tool supports optional user input to validate $\Delta$ and to specify custom
TFDs, thereby addressing \textbf{C4}.

\smallskip \noindent\emph{{Detecting violating tuples.}} \tool flags tuples
within the inferred periodicity that violate TFDs. E.g., in
Figure~\ref{fig:data_snippet}, $\mathtt{a_1}$, $\mathtt{a_3}$, and
$\mathtt{a_4}$ violate $TFD_1$ due to multiple \texttt{\small RSRP}, while
$\mathtt{a_7}$ due to missing \texttt{\small RSRP}.

% \af{This is good, but start with it saying that without any user input things can be very difficult at times. To remedy this \tool solicits user input when X, Y, Z happen.}

\subsubsection{\textbf{DSL for WN}} \label{dsl}
To address \textbf{C2}, we need a domain-specific language that includes common
operators used in WN data analysis. In this demo, we include frequently used
operators identified through analyzing scripts from members of the
POWDER~\cite{powder} team.

\definecolor{codebg}{gray}{0.94}

    \lstset{
      language=Python,
      basicstyle=\scriptsize\ttfamily,
      breaklines=true,
      backgroundcolor=\color{codebg},
      frame=single,
      rulecolor=\color{gray!50},
      xleftmargin=2pt,
      xrightmargin=2pt,
      aboveskip=1pt,
      belowskip=0pt,
      showstringspaces=false,
      tabsize=2,
      keywordstyle=\color{blue},
      commentstyle=\color{gray!70},
      stringstyle=\color{green!50!black}
    }

\begin{enumerate}[leftmargin=*]
    \item \textbf{Upsample}: Achieves the given temporal granularity by
    inserting new rows and filling in empty cells (Step 3 of
    Example~\ref{ex:one}). To do this, the semantic profile is used to choose between the following options: (a) forward filling, (b) backward filling, (c) interpolation, using a combination of the grouping attributes found by the semantic profiler to group and impute newly created rows .

    \begin{lstlisting}
        A["RSRP"] = A.groupby("frequency")["RSRP"].ffill()
    \end{lstlisting}
    % A["RSRP"] = A.groupby("frequency")["RSRP"].ffill()
	
    \item \textbf{Downsample}: Achieves the given temporal granularity by
    aggregating existing rows (Step 2 of Example~\ref{ex:one}). To do this, a combination of the grouping attributes found by the semantic profiler is used to group and aggregate using functions (e.g., logarithmic mean) from a pre-defined list of aggregation functions. An example of a downsample call would translate to the following \verb|pandas| code:

    \begin{lstlisting}
        df.groupby(["timestamp", "frequency"]).agg({"throughput" : "mean"})
    \end{lstlisting}
	
    \item \textbf{Impute}: Imputes a cell using the given technique such as
    forward-fill or backward-fill (Step 1 of Example~\ref{ex:one}), using a combination of the grouping attributes found by the semantic profiler to group and impute. Translates to a \verb|pandas| forward/backward fill operation. Example:

    \begin{lstlisting}
        df["latitude"] = df["latitude"].ffill()
    \end{lstlisting}

    \item \textbf{Rounds}: Rounds differently scaled units to the nearest homogeneous unit (e.g., rounding millisecond timestamps to seconds, based on the specified periodicity (Step~2 of Example~\ref{ex:one})). Translates to a \verb|pandas| code that is crafted based on the semantic type of the column. For example, for a time column:

    \begin{lstlisting}
        df["timestamp"] = df["timestamp"].dt.floor("S")
    \end{lstlisting}

    \item \textbf{Cast}: Cast from one unit to another unit to facilitate joins over columns with different representations of the same entity (Example~\ref{ex:two}). As discussed previously, the current version of \tool allows for type casting for WN specific types. Some examples:
    \begin{itemize}
        \item EARFCN $\xleftrightarrow{} $MHz (for frequency channels)
        \item Integer $\xleftrightarrow{}$ Hexadecimal (for cell IDs)
        \item Object $\xrightarrow{}$ Datetime (for timestamp columns)
        \item Object $\xrightarrow{}$ Float (for numerical columns)
    \end{itemize}
  
    \item \textbf{Drop row}: Drops rows based on the given conditions. For instance, drop rows that are not numerical in a column that is profiled to be numerical.

\end{enumerate}

\noindent\emph{Remark:} The above DSL is specific for handling WN-data issues
and should be treated as insufficient to resolve generic issues such as formatting
inconsistencies (e.g., date format), for which a generic DW tool~\cite{cowrangler, auto-suggest,
autodwts} can be used.

\subsubsection{\textbf{Scoring method}} 
\label{soloverview_ranking}

\tool generates a set of candidate operations by parameterizing the DSL
operators with appropriate parameters. During this step, \tool prunes
semantically invalid candidates, i.e., those that violate WN-specific
semantics, such as using an arithmetic mean as an imputation technique for
\texttt{\small RSRP} by querying an LLM to verify which of the generated DSL instances are accurate. This automates away the filtering a domain expert would encounter if all options were provided to only present semantically valid wrangling operations. An example when querying GPT-5.2 is provided below:

\begin{tikzpicture}
\node[
    draw,
    rounded corners,
    fill=pink!10,
    text width=8cm,
    align=left
] 
{\textbf{Question}:\\
You are an expert wireless data profiler. \\
I have a column in a pandas dataframe named RSRP. \\
It has a unit of dBm. \\
Here are some sample values:\\
\dots \\
You should decide what aggregation method to use to combine multiple values into one data point.\\
Your choices are: ['mean', 'sum', 'mode', 'median', 'log\_mean'].  \\
Their descriptions are: ['arithmetic mean', 'sum', 'most frequent value', '50th percentile value', 'convert to linear domain, do arithmetic mean, convert back to log domain']. \\
Answer only one of ['mean', 'sum', 'mode', 'median', 'log\_mean'] and nothing else.\\
\textbf{Answer}:\\
log\_mean
};
\end{tikzpicture}

While this reduces the search space, a key challenge
(\textbf{C3}) remains: determining which of these candidates should be
suggested to the user. To this end, \tool employs an efficient scoring method
that estimates the expected improvement in data quality---i.e., the reduction
in constraint violations---if a candidate wrangling operation were applied to
the data.

Given a dataset $D$, discovered constraints or rules $R$, semantic profiles $S$,
candidate wrangling operations $W$, and a violation function $V(D, R)$ that
computes the degree of violation by $D$ w.r.t $R$, the scoring method applies
each $w \in W$ to a small sample of the data $D'$ to obtain $D_w'$, where the sampling technique
used is aware of the WN-specific semantics. This enables \tool to efficiently
estimate the expected data-quality improvement when $w$ is applied over $D$ as
$V(D', R) - V(D_w', R)$. While minimizing constraint violations is the primary
objective, \tool also accounts for data side effects by ensuring that each of
the $k$ suggested wrangling operations satisfy a predefined data-side-effect
budget (e.g., at most $p$\% cells/rows of the data can be modified).

%%% DEMO FIGURE. Place here to make sure it appears top of the page
\begin{figure*}[t]
\resizebox{0.77\textwidth}{!}
{
	\includegraphics[]{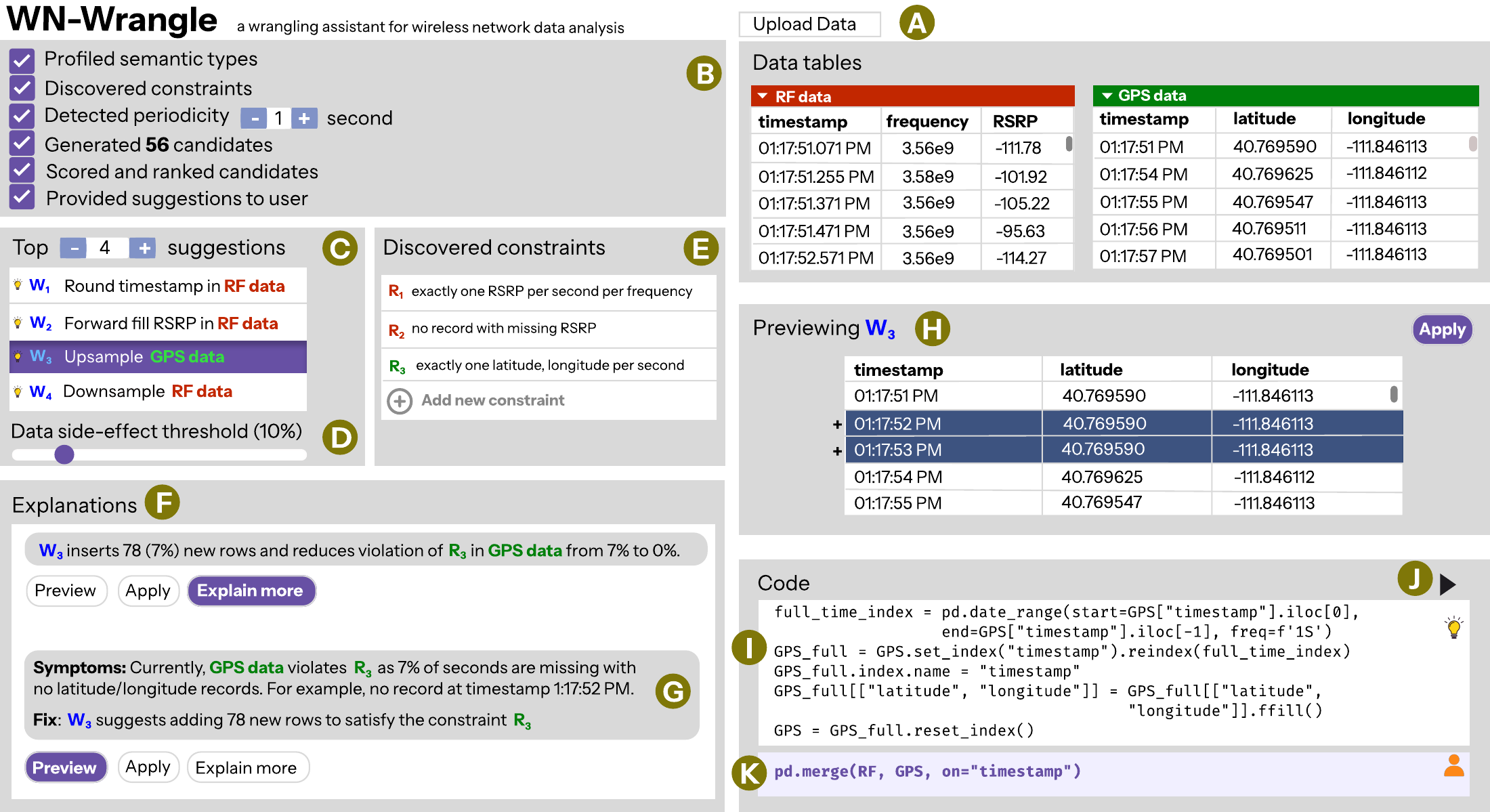}
}
\vspace{-2.5mm}

\caption{\small \tool interface: 
\stepCounter{A} data upload and preview; 
\stepCounter{B} progress tracker for the \tool workflow; 
\stepCounter{C} suggested wrangling operations; 
\stepCounter{D} user-specified threshold on data side effects; 
\stepCounter{E} discovered constraints; 
\stepCounter{F} explanations of the suggestions with interactive support; 
\stepCounter{G} follow-up clarifications; 
\stepCounter{H} on-demand preview of a selected suggestion; 
\stepCounter{I} editable code synthesized by \tool for the selected suggestion; 
\stepCounter{J} execution button to apply the suggestion to the full dataset; 
\stepCounter{K} custom user code for joining the wrangled tables.}

\vspace{-3mm}

% \ak{Figma link: https://www.figma.com/design/IJcidRPyVcVmuQ6ahnEDAP/RF-Wrangle?node-id=0-1\&t=9EIwvF0LBlkPXIyr-1}

\label{fig:demo}
\end{figure*}

\subsubsection{\textbf{Explanation module}}
\label{explanation_module}
Finally, the explanation module combines the information from the semantic
profiler, constraints discovered and violations detected by the constraint
discovery module, and the estimated reduction in violation obtained by the
scoring method to generate a human-understandable, natural-language explanation
for each of the suggested wrangling operations.

% \af{This is a good point, but I don't think we will have space for it. Also,
% why does it appear in the TFD/constraints part? It is talking about
% appropriate imputation strategy, which you talk about in Semantic profiler.}
% For example, in GPS measurements, users may not desire forward-filling as they
% notice that forward-filling GPS measurements may lead to sudden jumps in GPS
% positions, especially where the GPS connections might have been lost. If the
% user knew they drove a maximum of 30 miles per hour, then it would be
% impossible for two consecutive seconds of GPS measurements to have a distance
% larger than $8.3 \times 10^{-3}$, hence, they can manually also provide such a
% constraint in the form: \verb|[|$\Delta = \verb|1 second|\verb|]|
% \xrightarrow[]{} \verb|[distance| \leq \verb|8.3e-3]|$. In this case, missing
% seconds of GPS locations would instead by filled with interpolation instead to
% satisfy the constraint.

%!TEX root = main.tex

\section{Demonstration} \label{demo}

We will demonstrate \tool over real-world POWDER datasets~\cite{powder_dataset}. We will guide users through eleven steps of Figure~\ref{fig:demo} impersonating Naomi over the dataset~\cite{dataset_link} of Example~\ref{ex:one}.

\smallskip Step~\stepCounter{A}~\textbf{(Data upload and preview)}. The user uploads two data files \texttt{\small RF.csv} and \texttt{\small GPS.csv} and previews the data. \tool displays the first five rows. The user can scroll to see more.

\looseness-1 Step~\stepCounter{B}~\textbf{(Workflow progress tracker)}. \tool semantically profiles the data attributes (\S\ref{semantic_profiler}); discovers temporal and other constraints and detects the periodicity parameter $\Delta$  (\S\ref{rule_discovery}), which the user can refine if they wish to; generates 56 wrangling candidates, scores, and ranks them (\S\ref{soloverview_ranking}) to generate the suggestions.

Step~\stepCounter{C}~\textbf{(Wrangling suggestions)}. \tool suggests $4$ operations: $W_1$, $W_2$, \& $W_4$ for the \texttt{\small RF} data and $W_3$ for the \texttt{\small GPS} data.

Step~\stepCounter{D}~\textbf{(Data side-effect)}. 
Along with the suggestions, \tool displays maximum data side-effect incurred (10\% rows were impacted) by any of the suggestions. The user can adjust the side-effect threshold and \tool ensures that each suggested operation satisfies the user-specified side-effect requirement.

Step~\stepCounter{E}~\textbf{(Discovered constraints)}. 
\tool lists the constraints (three in this case) that it used to compute the degree of violation for scoring the candidate wrangling operations. For example, the constraint $R_1$: ``\textit{exactly one \texttt{\small RSRP} record per second per frequency}'' applies to the \texttt{\small RF} data. The user can add new constraints.

Steps~\stepCounter{F} \& \stepCounter{G}~\textbf{(Explanation and interaction)}.
The user wants to understand the rationale behind the suggestion $W_3$.
\tool provides an initial explanation---``\textit{$W_3$ inserts 7\% new rows,
satisfying $R_3$}''---along with options to \emph{Preview} its impact,
\emph{Apply} $W_3$ to the full data, or request further
\emph{Explanation}. After selecting ``\emph{Explain more},'' \tool offers a
detailed explanation describing missing readings across multiple seconds.
Satisfied, the user chooses to \emph{Preview} the impact of $W_3$ on the
\texttt{\small GPS} data.

\looseness-1 Step~\stepCounter{H}~\textbf{(Preview suggestion impact)}. \tool previews the impact of $W_3$ on a small sample of the \texttt{\small GPS} data, highlighting newly inserted rows by ``+''. Satisfied, the user clicks on ``Apply''.

Steps~\stepCounter{I} \& ~\stepCounter{J}~\textbf{(Automatically generated wrangling code)}. \tool inserts an editable code snippet for $W_3$ to the user notebook, and automatically executes it in Step~\stepCounter{J}. The user accepts the other three suggestions in a similar way (not shown).

Step~\stepCounter{K}~\textbf{(Custom code)}. Finally, the user successfully joins the two (now wrangled) datasets as per Example~\ref{ex:one}. 

\smallskip While \tool applies to any WN data scenario, it is especially useful for experimental wireless testbeds like POWDER~\cite{powder}, where hundreds of users generate diverse datasets from $2{,}000+$ yearly experiments, supporting advanced wireless applications.

% \smallskip We envision that \tool will enable hundreds of users on the POWDER platform~\cite{powder} who run over $2000$ experiments per year to produce high quality wireless-network datasets for enhanced wireless applications and use cases. 
% We aim to extend \tool to support data wrangling over unstructured data (e.g., wireless application logs,) wrangling to enable augmentation with non-wireless data (e.g., relating WN data to climate patterns,) wireless downstream task aware wrangling, and using fine-tuned LLM wrangling experts that can potentially improve accessibility to a variety of stakeholders. 

% With the joined dataset, participants can view the measurements from the
% dataset on a Google Earth interface. Moreover, we will also engage participants
% in reviewing the improvement in performance of a time-series forecasting ML
% model trained on other datasets from the POWDER platform, with/without the
% re-sampling steps used in the demonstration. This demonstration outlines how WN
% datasets require special wrangling techniques to facilitate the generation of
% high-quality, analysis-ready datasets. Extending this work can bring about new
% wrangling paradigms, such as wrangling unstructured WN data (e.g., from
% logs) discovering data augmentation choices plans with non-WN data (e.g.,
% relating WN data with weather) fine-tuned LLM wrangling experts, and
% supporting more custom DSL operations.

% \af{Mention how this work can be extended for other domains such as healthcare
% where multiple devices may measure patient data.}

\section{Motivating Results} \label{results}

% We will demonstrate \tool over real-world POWDER datasets~\cite{powder_dataset}. We will guide users through eleven steps of Figure~\ref{fig:demo} impersonating Naomi over the dataset~\cite{dataset_link} of Example~\ref{ex:one}.
In this section, we discuss some preliminary results of how the wrangling suggested by \tool improves the performance of regression models on the dataset described in Example~\ref{ex:one}.

\smallskip \noindent We trained 3 basic regression models (Linear Regression, Lasso, and Random Forest) on two versions of the dataset. The ``Before'' data version in Table~\ref{tab:results} refers to the merged dataset that was not down/up sampled, and one which only had rounded timestamps to ensure some rows could be merged (all rows are not included). The ``After'' version refers to the final dataset produced by \tool.
\vspace{-3mm}
\begin{table}[hb]% h asks to places the floating element [h]ere.
  \caption{RMSE values of regression models on merged dataset before and after wrangling by \tool.}
  \label{tab:results}
  \begin{tabular}{cccc}
    \toprule
    Data version & Linear Regression & Lasso & Random Forest\\
    \midrule
    Original & $8.15$ dBm & $8.2$ dBm & \textbf{4.14 dBm} \\
    Wrangled & \textbf{7.69 dBm} & \textbf{7.69 dBm} & $4.23$ dBm \\
  \bottomrule
\end{tabular}
\end{table}

\vspace{-4mm}
As evidenced in the results Table~\ref{tab:results}, the effect of accurate wrangling steps suggested by \tool induces a $0.46$/$0.51$ dBm improvement in predicting the signal strength for the Linear Regression/Lasso models only using the location and frequency channels as features during prediction. This can be attributed to the inclusion of more training data rows (that are not dropped, rather forward filled) that the model can use to better learn \texttt{\small RSRP} predictions. For the Random Forest model, we notice diminishing returns, as it learns the concept reasonably well. However, it is clear that the lack of domain-specific wrangling steps can lead to unexpected loss in performance for joined WN datasets, depending on the magnitude of missing data.

\bibliographystyle{ACM-Reference-Format}
\bibliography{references}

\end{document}